\documentclass[conference]{IEEEtran}
\IEEEoverridecommandlockouts
\usepackage{cite}
\usepackage{amsmath,amssymb,amsfonts}
\usepackage{algorithmic}
\usepackage{textcomp}
\usepackage{caption}
\usepackage{graphicx}
\usepackage{pgfplots}
\pgfplotsset{compat=1.18}
\usepackage{subcaption}
\usepackage{multirow}
\usepackage{algorithm}
\usepackage{makecell}
\usepackage{array}
\usepackage{authblk}

\usepackage{xcolor}
\makeatletter
\newcommand{\thickhline}{%
    \noalign {\ifnum 0=`}\fi \hrule height 1pt
    \futurelet \reserved@a \@xhline
}
\newcolumntype{"}{@{\hskip\tabcolsep\vrule width 1pt\hskip\tabcolsep}}
\makeatother
\def\BibTeX{{\rm B\kern-.05em{\sc i\kern-.025em b}\kern-.08em
    T\kern-.1667em\lower.7ex\hbox{E}\kern-.125emX}}
\usepackage{comment}
\usepackage{fancyhdr}

\begin{document}

\title{Exploration of Activation Fault Reliability in Quantized Systolic Array-Based DNN Accelerators}

\author{
{Mahdi Taheri}\textsuperscript{1},\
{Natalia Cherezova}\textsuperscript{1},\
{Mohammad Saeed Ansari}\textsuperscript{2},\
{ Maksim Jenihhin}\textsuperscript{1},\\
{Ali Mahani}\textsuperscript{3,4},\
{Masoud Daneshtalab}\textsuperscript{1,5},\
{Jaan Raik}}

\affil[1]{Tallinn University of Technology, Tallinn, Estonia}
\affil[2]{University of Alberta, Edmonton, Canada,}

\affil[3]{Shahid Bahonar University of Kerman, Kerman, Iran}
\affil[4]{York University, Toronto, Canada}
\affil[5]{Mälardalen University, Västerås, Sweden}
\affil[1]{mahdi.taheri@taltech.ee}

\maketitle

\begin{abstract}
The stringent requirements for the Deep Neural Networks (DNNs) accelerator's reliability stand along with the need for reducing the computational burden on the hardware platforms, i.e. reducing the energy consumption and execution time as well as increasing the efficiency of DNN accelerators. Moreover, the growing demand for specialized DNN accelerators with tailored requirements, particularly for safety-critical applications, necessitates a comprehensive design space exploration to enable the development of efficient and robust accelerators that meet those requirements. Therefore, the trade-off between hardware performance, i.e. area and delay, and the reliability of the DNN accelerator implementation becomes critical and requires tools for analysis.
This paper presents a comprehensive methodology for exploring and enabling a holistic assessment of the trilateral impact of quantization on model accuracy, activation fault reliability, and hardware efficiency. A fully automated framework is introduced that is capable of applying various quantization-aware techniques, fault injection, and hardware implementation, thus enabling the measurement of hardware parameters. Moreover, this paper proposes a novel lightweight protection technique integrated within the framework to ensure the dependable deployment of the final systolic-array-based FPGA implementation. The experiments on established benchmarks demonstrate the analysis flow and the profound implications of quantization on reliability, hardware performance, and network accuracy, particularly concerning the transient faults in the network's activations.
\end{abstract}

\begin{IEEEkeywords}
deep neural networks, design space exploration, quantization, fault simulation, reliability assessment
\end{IEEEkeywords}

\section{Introduction}
In the past decades, Deep Neural Networks (DNNs) demonstrated a significant improvement in accuracy by adopting intense parameterized models \cite{taheri2022dnn}.
As a consequence, the size of these models has drastically increased, imposing challenges in deploying them on resource-constrained platforms \cite{gholami2021survey}. 
FPGAs are a widely used solution for flexible and efficient DNN accelerator implementations and have shown superior hardware performance in terms of latency and power \cite{riazati2022autodeephls}. 
In practice, deployment of an FPGA-based DNN accelerator for the safety- and mission-critical applications (e.g., autonomous driving) requires addressing the trade-off between different design parameters of \textit{hardware performance}, e.g., area, power, delay, and \textit{reliability}.
\begin{figure}[t]
    \includegraphics[width=0.5\textwidth]{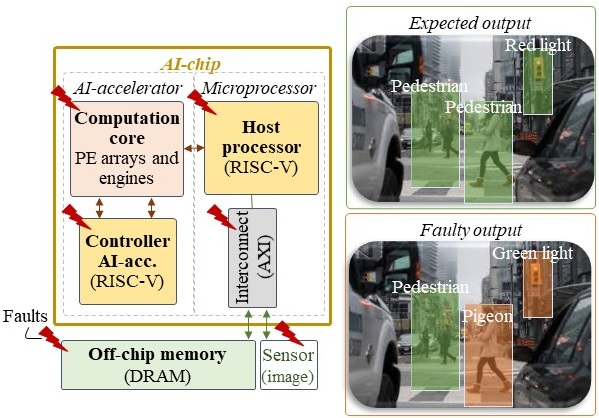}
    \centering
    \caption {Hardware-induced reliability threats in an example DNN accelerator and their possible impact on the output}
    \label{fig:rel_thr}
\end{figure}
A compromise between conflicting requirements can be achieved by simplifying the implementation to sacrifice the precision of results but benefiting from lower resource utilization, energy consumption, and higher system efficiency. \textit{Quantization} is one such concept that is being widely used in neural network deployments \cite{taheri2023deepaxe}. Quantization is used to compress the model for storage and computation reduction. However, recent research shows that faults in memory can cause a significant drop in DNN accuracy, which raises concern about the impact of quantization on the reliability of the network \cite{choi2023rq}.

The reliability of DNN accelerators expresses their ability to produce correct outputs in the presence of hardware faults originating from various phenomena, e.g., radiation-induced soft errors in memory or logic \cite{ahmadilivani2023systematic}.
DNNs are known to be inherently fault-resilient due to the high number of learning process iterations and several parallel neurons with multiple computation units. Nevertheless, faults may impact the output accuracy of DNNs drastically \cite{taheri2023noise}, and in the case of resource-constrained critical applications, the reliability of DNNs is required to be evaluated and guaranteed \cite{siddique2021exploring}. 
The complexity of such evaluation motivates an \textit{automated toolchain} with quantization and reliability analysis to support \textit{Design Space Exploration (DSE)} for DNN accelerators already at the early design stage, i.e. starting from a high-level description, followed by providing an FPGA prototype for the selected design.

While the protection of weights stored in ROM can be ensured through error correction codes (ECC) or similar protection techniques, the dynamic nature of activations, which are stored for a short period of time in usually unprotected memories, poses a critical concern. Thus, it is crucial to thoroughly investigate the consequences of faults in the network's activations.

This paper presents a framework containing a fully automated toolchain to perform a study on the impact of quantization on network accuracy, hardware performance, and reliability drop in the presence of activation faults (Fig. \ref{fig:rel_thr}) in systolic-array-based FPGA accelerators. To the best of our knowledge, this is the first framework that holistically considers those parameters.
A novel lightweight mitigation technique is proposed and integrated into the framework to study potential trade-offs of compensating the reliability drops. 
The proposed methodology enables the analysis both at the level of the network model and at the level of individual layers of the network.

This framework is empowered by techniques for quantizing the networks and restricting the activation ranges to be limited to a certain level throughout the whole network execution by applying an extra scaling function in the network inference. This framework uses the high-level description of a DNN as an input and is capable of providing a transient-fault-resilient systolic-array-based FPGA implementation of the network utilizing the design parameters selected by the DSE. The main contributions in this work are as follows:

\begin{itemize}
    \item A methodology for holistic exploration of quantization and reliability trade-offs in systolic-array implementation that enables assessing the trilateral impact of quantization on accuracy, activation fault reliability, and hardware performance. 
    \item A fully-automated framework that is capable of applying quantization-aware training, post-training quantization, range-restriction, fault simulation, and implementing the whole methodology down to hardware implementation to measure actual hardware parameters like area, latency, etc.
    \item A lightweight and effective protection technique is developed and adopted in the framework toolchain to provide the final reliable systolic-array-based FPGA implementation of the network
    \item Demonstration and analysis of the results on the impact of quantization on reliability, hardware performance, and accuracy of the neural networks due to the transient faults in the activations for two well-known benchmarks.

\end{itemize}


The rest of the paper is organized as follows. Related works are discussed in Section II, the methodology and framework are presented in Section III, the experimental setup and results are provided in Section IV, and finally, the work is concluded in Section V.

\section{Related Works}
\subsection{DNN reliability and quantization studies}
Several works examine the impact of different fault models on the basis of a number of layers in DNNs and different data types \cite{li2017understanding}. Investigation into the effects of data precision is done in \cite{basso2020impact}, where authors conducted a comparison of the resilience of FP16, FP32, and FP64 in the context of Matrix Multiplication. Their findings indicated that the reduction of precision not only enhances GPU performance and efficiency but also contributes to its overall resilience.

Another study \cite{libano2021reduced} involved the deployment of MNIST CNN on FPGAs utilizing FP32, FP16. The results of the experiment demonstrated that decreasing the data precision in CNNs can lead to a substantial enhancement in overall resilience. This improvement was attributed to the reduced memory usage. Furthermore,\cite{libano2020understanding} noted that the application of binary quantization to weights in convolutional layers results in decreased vulnerability factors, although it does increase the criticality of faults.
\cite{syed2021fault} showed that the impact of faults is higher in most significant bits (MSBs) and with more aggressive compression the most significant bits are more probable to be exposed to faults. The aforementioned works show that quantization from higher data representations like FP32 down to INT16 has a positive impact on the performance and overall resilience, though \textit{on the lower quantization ranks this matter should be studied and is not always impacting positively on the resilience}

In \cite{arechiga2018effect}, it is shown that in some cases, the impact of the faults in the weight memories of a DNN can be negligible. Even though in the above-mentioned works, \textit{impact of faults (soft errors modeled as bit flips) in the weights of a DNN during inference is examined}, to further enhance our comprehension of the impact of quantization on the reliability of DNNs in systolic-array-based DNN accelerators, this work is enriched with an FI engine capable of injecting faults into the activations of the DNNs in the systolic architecture.
 \begin{figure*}[h]
    \centering
    \includegraphics[width = 0.9\textwidth]{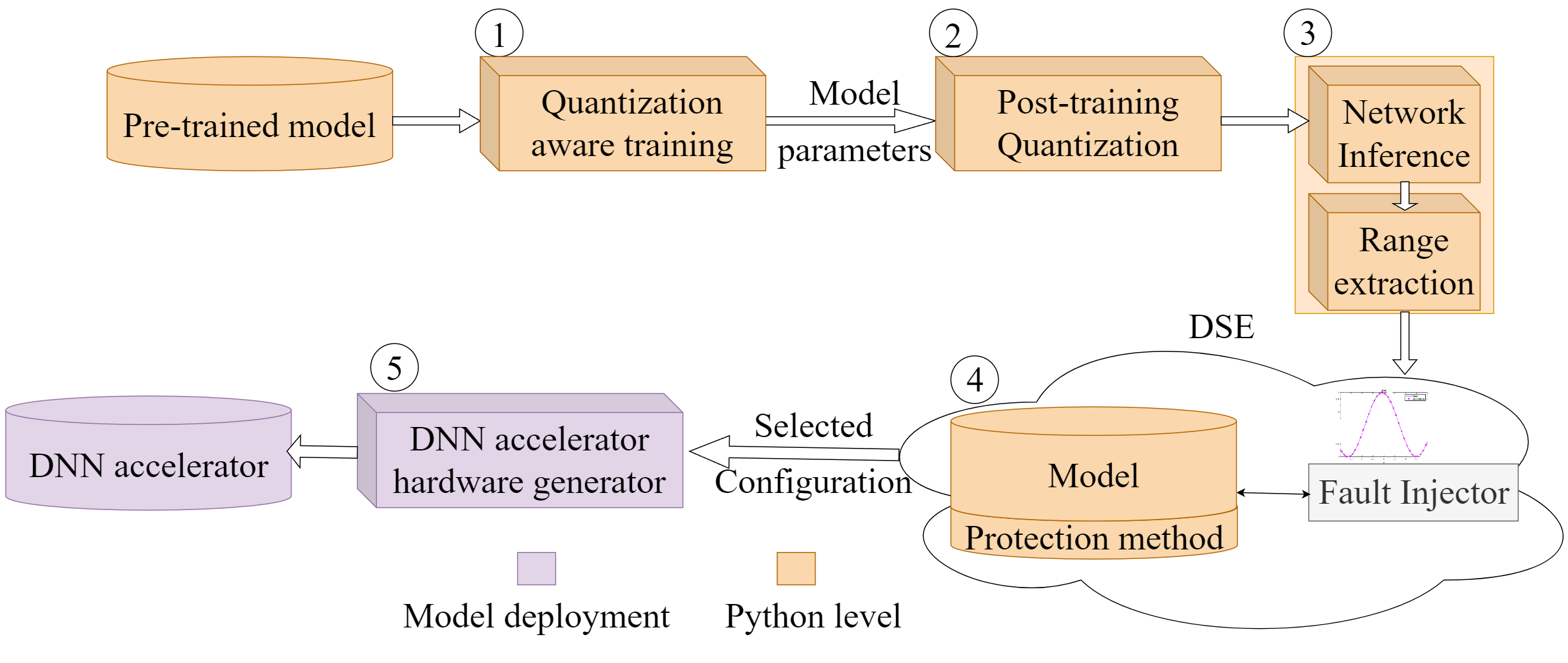}
    \caption{Proposed methodology flow}
    \label{method}
\end{figure*}
\subsection{Fault mitigation techniques}
The process of quantization and outlier regularization offers the potential to restrict the numerical range within a DNN, thereby eliminating the possibility of generating excessively large values due to faults \cite{choi2023rq}. 


Hoang et al. analyzed how various boundary values affect the network's accuracy. They have found that the best boundary values for each layer are not necessarily the maximum values of the layers' activations \cite{hoang2020ft}. Hence, they propose an interval search algorithm to find appropriate boundary values for the ReLU activation function at each layer, named FT-ClipAct. The proposed clipped activation function maps their outputs to 0 if activations exceed the boundaries.
Although these methods can decrease the effect of faults in DNNs, they remove a significant portion of non-zero activations by replacing them with zero, leading to an accuracy drop in high error rates. It is also noteworthy that the mentioned methods do not consider low integer quantization and are mostly working with FP32 and FP16. In this paper, we introduce a novel \textit{lightweight range-checking} circuit that, despite the other works, can consider the maximum values of the layers' activations and replace the out-ranged values with either lower- or upper-bound to avoid fault propagation and also avoid removing a significant portion of non-zero activations by replacing them with only zero. This protection technique is employed in the DNN accelerator hardware generation step of the framework to provide the user with a prototype of the reliable accelerator.

\subsection{DNN hardware accelerator frameworks}
The advantages of implementing and deploying DNNs on FPGAs are advocated in several recent works. The existing FPGA-based toolchains to map Convolutional Neural Networks (CNNs) are presented in the surveys \cite{venieris2018toolflows, guo2019dl, abdelouahab2018accelerating, molina2022high}. The FINN framework \cite{FINN} is released by Xilinx for the exploration of quantized CNNs' inference on FPGAs that also provide customized data-flow architectures for each network. 
Heterogeneous systems are another design strategy in the automated toolchains that propose hardware/software co-design \cite{1, ghaffari2020cnn2gate, mousouliotis2020cnn}. In these designs, computational units, e.g., adders and multipliers, are mainly implemented on Programmable Logic (PL) that is controlled by a control unit in a CPU using a dedicated framework, e.g., OpenCL \cite{stone2010opencl}. In this work, we introduce a hardware generation step as part of the framework, to explore DNN inference on an FPGA-based accelerator with a customizable systolic array.
It seamlessly integrates with the PYNQ framework\cite{wang2018pynq}, leveraging the original PYNQ bootable image. This integration enhances versatility and compatibility, enabling users to implement their network on different FPGA devices supporting PYNQ. Furthermore, the reconfigurable systolic array implementation introduced in this step provides flexibility and scalability. Users can customize this step to meet their specific network requirements by providing trained parameters and network architectures, resulting in efficient and high-performance DNN inference.

To the best of our knowledge, none of the previous works explored the impact of using different levels of full quantization (weights, activations and biases) of a DNN in the presence of transient faults in the activations on the reliability, accuracy, and delay/resource utilization of the target DNN accelerator.

The approach proposed in this paper goes beyond the state of the art by establishing a fully automated tool for enabling efficient quantization in FPGA-based DNN accelerators aimed at safety-critical applications. The proposed framework contains a high-level simulator to study the impact of quantization on the reliability and accuracy of the network by considering the hardware architecture, with and without protection techniques, followed by an efficient and user-friendly heterogeneous FPGA implementation of the selected DNN configuration. 

\section{Proposed Methodology}

Fig. \ref{method} illustrates the methodology flow established in the toolchain for reliability and hardware performance analysis of quantized DNN hardware accelerators. This framework takes the DNNs' \emph{Pre-trained model} description as the input. The design, training, and testing of the DNNs are performed in Python. \emph{Quantization-aware training} and the \emph{Post-training quantization}, \emph{Range extraction} and \emph{DSE} steps are seamlessly integrated into the same environment and are responsible for extracting the required data for the hardware generation step. This step is responsible for the hardware implementation of the selected configuration to measure actual hardware parameters like area, latency, etc.

\textbf{Step 1: Quantization-aware training.}
For this purpose, a full quantization is implemented, targeting all activations, weights, and biases. The framework first takes the description of the network provided by the user and then uses the TFlite library for quantization-aware training. The user can replace their preferred quantization library with the toolchain for this step. The main output of this step is the quantized network's parameters (weights and biases) and network architecture.

\textbf{Step 2: Post-training quantization.} In the post-training quantization step, the user can define any further quantization that can be applied to the network with a negligible accuracy loss depending on the level of the quantization. This framework supports quantizing the network down to 4-bit INT. The output accuracy of the generated network is also provided at this step and is kept as a baseline for the further steps of the methodology. For this step, the following algorithm is applied to the network parameters:

The mapping equation is defined as:
\[
\tilde{x} = \text{clamp} \left( \left\lfloor \frac{x}{S} \right\rfloor + Z; q_{\text{min}}, q_{\text{max}} \right)
\]

\[
S = \frac{x_{\text{max}} - x_{\text{min}}}{2^b - 1}
\]

Where Z is the offset defined as zero-point, $x_{\text{max}}$ and $x_{\text{min}}$ represent the maximum and the minimum value in the vector. The quantization range $[q_{\text{min}}, q_{\text{max}}]$ is determined by the bit-width. We focus solely on uniform unsigned symmetric quantization, as it is the most commonly employed quantization setup. Hence, $q_{\text{min}}$ is equal to 0, and $q_{\text{max}}$ is equal to $2^b-1$, where $b$ denotes the bit-width, determining the number of integer grids.

\textbf{Step 3: Inference and range extraction.}
In this step, after running the inference, the ranges of the activations are extracted for evaluation and reliability study. The ranges are extracted based on the set of validation data, and then the framework extracts the next set of ranges for each layer based on the test data and validates the extracted data correspondingly.

\textbf{Step 4: Design Space Exploration.}

\textbf{Step 4-A: Fault simulation.} Reliability analysis relies on a Fault Injection (FI) in a \textit{systolic-array-based simulation} of the network in Python, assuming the single bit-flip faults in the activations. While the multiple-bit fault model is more accurate, it requires a prohibitively large number of fault combinations to be considered. Fortunately, it has been shown that high fault coverage obtained using the single-bit model results in a high fault coverage of multiple-bit faults \cite{bushnell2004essentials}. Therefore, a vast majority of practical FI and test methods are based on the single-bit fault assumption.
However, this framework is capable of applying multiple-bit-flips as a fault model depending on the user demand.

The reliability analysis step applies the accuracy drop comparison of the network-under-test as one of the assessment metrics.
In addition, the framework assesses the reliability of the DNN by comparing the output probability vector of the golden run (i.e. the DNN that behaves as expected, without faults) and the faulty run (i.e. the DNN that includes the fault).
These metrics involve the SDC (Silent Data Corruption) rate. Specifically, one of the two metrics is ``absolute'', and the other one is ``relative''. The SDC rate is defined as the proportion of faults that caused misclassification in comparison with the golden model.
\begin{itemize}
	\item \textbf{SDC-1:} Fault caused a misclassification in the top-ranked output class.
	\item \textbf{SDC-5:} Fault caused the top-ranked element not to exist in the top-5 predicted output classes.
	\item \textbf{SDC-10\%:} Fault caused a variation in the output confidence score of the top-ranked output class more than 10\% compared to the golden model.
\end{itemize}

After choosing the preferred quantization in \textbf{Step~2}, the designer can go through the systolic-array-based fault injector provided for the reliability evaluation of the Quantized DNN (QDNN). The final design is fed to the next step \emph{hardware generator} for the DNN hardware accelerator generation and hardware performance evaluation process. 
\begin{figure}[h]
    \centering
    \includegraphics[width=0.35\paperwidth]{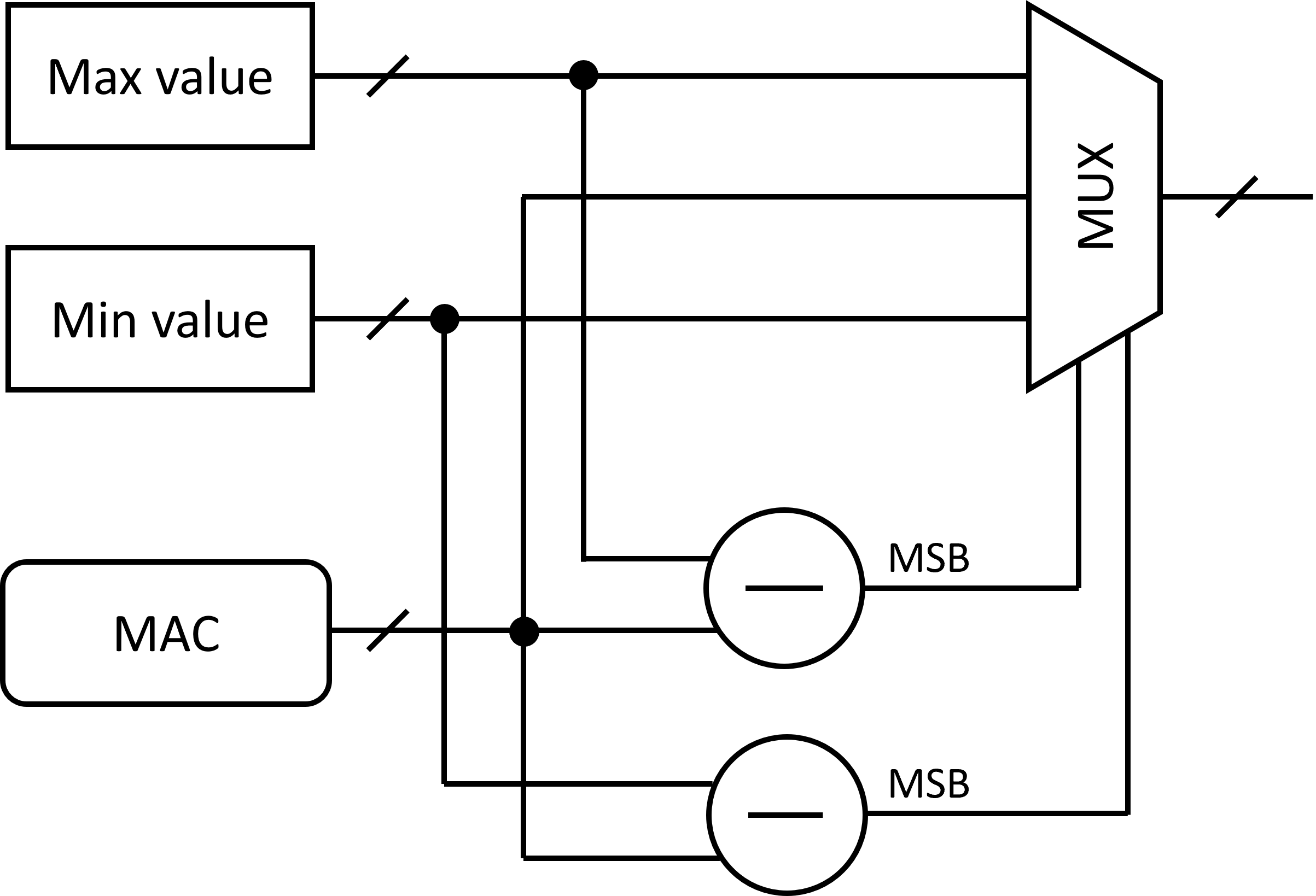}
    \caption{Proposed lightweight mitigation technique}
    \label{mit}
\end{figure}
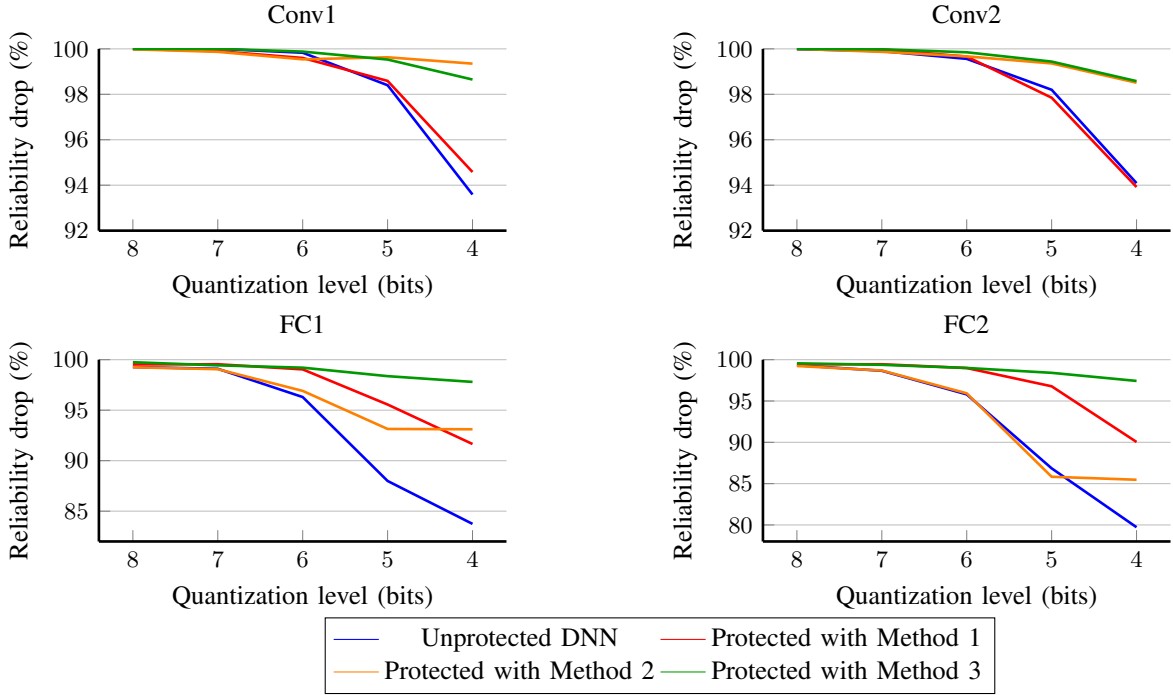
\begin{figure*}[!h]
\begin{center}
\begin{subfigure}{.48\textwidth}
\begin{tikzpicture}
\begin{axis} [height=4cm, width=7cm, ymin=92, ymax=100, axis lines*=left,
              ymajorgrids=true, line width=1pt,
              x dir=reverse, xtick={8,7,6,5,4},
              enlarge x limits=0.1, tick label style={font=\small},
              xlabel=Quantization level (bits), ylabel={Reliability drop (\%)}, ylabel near ticks,
              title={Conv1}]
\addplot[color=blue] coordinates {
	(8,100)
	(7,100)
	(6,99.83)
        (5,98.4)
	(4,93.59)
};
\addplot[color=red] coordinates {
	(8,100)
	(7,99.9)
	(6,99.6)
        (5,98.59)
	(4,94.58)
};
\addplot[color=orange] coordinates {
	(8,99.98)
	(7,99.87)
	(6,99.54)
        (5,99.63)
	(4,99.35)
};
\addplot[color=green!60!black] coordinates {
	(8,100)
	(7,100)
	(6,99.88)
        (5,99.53)
	(4,98.65)
};
\end{axis}
\end{tikzpicture}
\end{subfigure}
\begin{subfigure}{.48\textwidth}
\begin{tikzpicture}
\begin{axis} [height=4cm, width=7cm, ymin=92, ymax=100, axis lines*=left,
              ymajorgrids=true, line width=1pt,
              x dir=reverse, xtick={8,7,6,5,4},
              enlarge x limits=0.1, tick label style={font=\small},
              xlabel=Quantization level (bits), ylabel={Reliability drop (\%)}, ylabel near ticks, title={Conv2}]
\addplot[color=blue] coordinates {
	(8,99.99)
	(7,99.9)
	(6,99.56)
        (5,98.2)
	(4,94.09)
};
\addplot[color=red] coordinates {
	(8,100)
	(7,99.98)
	(6,99.66)
        (5,97.85)
	(4,93.92)
};
\addplot[color=orange] coordinates {
	(8,100)
	(7,99.87)
	(6,99.67)
        (5,99.36)
	(4,98.51)
};
\addplot[color=green!60!black] coordinates {
	(8,100)
	(7,99.98)
	(6,99.85)
        (5,99.44)
	(4,98.58)
};
\end{axis}
\end{tikzpicture}
\end{subfigure}
\begin{subfigure}{.48\textwidth}
\begin{tikzpicture}
\begin{axis} [height=4cm, width=7cm, ymin=82, ymax=100, axis lines*=left,
              ymajorgrids=true, line width=1pt,
              x dir=reverse, xtick={8,7,6,5,4},
              enlarge x limits=0.1, tick label style={font=\small},
              xlabel=Quantization level (bits), ylabel={Reliability drop (\%)}, ylabel near ticks, title={FC1}]
\addplot[color=blue] coordinates {
	(8,99.27)
	(7,99.1)
	(6,96.3)
        (5,87.99)
	(4,83.74)
};
\addplot[color=red] coordinates {
	(8,99.53)
	(7,99.55)
	(6,99.04)
        (5,95.56)
	(4,91.65)
};
\addplot[color=orange] coordinates {
	(8,99.25)
	(7,99.06)
	(6,96.91)
        (5,93.14)
	(4,93.11)
};
\addplot[color=green!60!black] coordinates {
	(8,99.75)
	(7,99.44)
	(6,99.21)
        (5,98.37)
	(4,97.8)
};
\end{axis}
\end{tikzpicture}
\end{subfigure}
\begin{subfigure}{.48\textwidth}
\begin{tikzpicture}
\begin{axis} [height=4cm, width=7cm, ymin=78, ymax=100, axis lines*=left,
              ymajorgrids=true, line width=1pt,
              x dir=reverse, xtick={8,7,6,5,4},
              enlarge x limits=0.1, tick label style={font=\small},
              xlabel=Quantization level (bits), ylabel={Reliability drop (\%)}, ylabel near ticks, title={FC2},
              legend columns=2, legend entries={Unprotected DNN,Protected with Method 1, Protected with Method 2, Protected with Method 3},
              legend to name=methods]
\addplot[color=blue] coordinates {
	(8,99.33)
	(7,98.66)
	(6,95.8)
        (5,86.85)
	(4,79.72)
};
\addplot[color=red] coordinates {
	(8,99.44)
	(7,99.45)
	(6,98.99)
        (5,96.78)
	(4,90.04)
};
\addplot[color=orange] coordinates {
	(8,99.24)
	(7,98.69)
	(6,95.93)
        (5,85.83)
	(4,85.47)
};
\addplot[color=green!60!black] coordinates {
	(8,99.56)
	(7,99.38)
	(6,99.00)
        (5,98.41)
	(4,97.43)
};
\end{axis}
\end{tikzpicture}
\end{subfigure}
\ref{methods}
\end{center}
\caption{Lenet-5 layer-level reports of reliability drop (based on FI for different quantized networks)}
\label{ax-accuracy}
\end{figure*}
\begin{figure}[!h]
\begin{center}
\begin{tikzpicture}
\begin{axis} [height=5cm, width=9cm, ymin=80, ymax=100, axis lines*=left,
              ymajorgrids=true, line width=1pt,
              x dir=reverse, xtick={8,7,6,5,4},
              enlarge x limits=0.1, tick label style={font=\small},
              xlabel=Quantization level (bits), ylabel={Reliability drop (\%)}, ylabel near ticks,
              legend style={font=\small,at={(0.5,-0.3)},anchor=north,legend columns=3}]
\addplot[color=blue] coordinates {
	(8,99.92)
	(7,99.99)
	(6,99.98)
        (5,99.98)
	(4,93.95)
};
\addplot[color=red] coordinates {
	(8,99.99)
	(7,99.99)
	(6,99.74)
        (5,98.75)
	(4,94.36)
};
\addplot[color=orange] coordinates {
	(8,99.9)
	(7,99.92)
	(6,99.78)
        (5,98.14)
	(4,93.97)
};
\addplot[color=green!60!black] coordinates {
	(8,99.91)
	(7,99.75)
	(6,99.42)
        (5,97.47)
	(4,93.78)
};
\addplot[color=cyan] coordinates {
	(8,99.88)
	(7,99.91)
	(6,98.77)
        (5,95.16)
	(4,87.66)
};
\addplot[color=gray] coordinates {
	(8,99.42)
	(7,98.79)
	(6,97.23)
        (5,91.96)
	(4,81.66)
};
\legend{Conv1,Conv2,Conv3,Conv4,Conv5,FC1}
\end{axis}
\end{tikzpicture}
\end{center}
\caption{AlexNet layer-level reports of reliability drop (\%) based on different quantization levels (unprotected design)}
\label{re_improv0}
\end{figure}
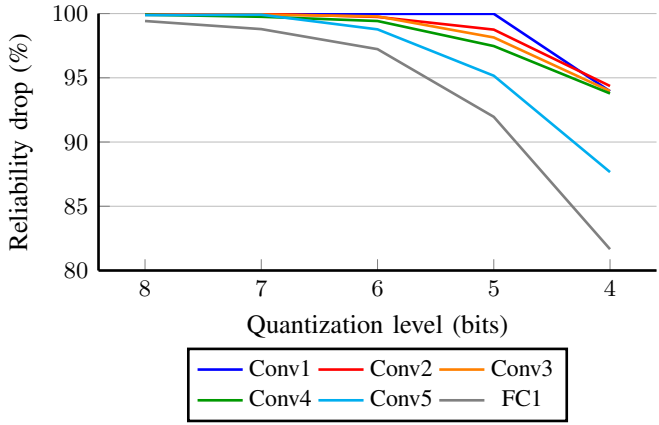

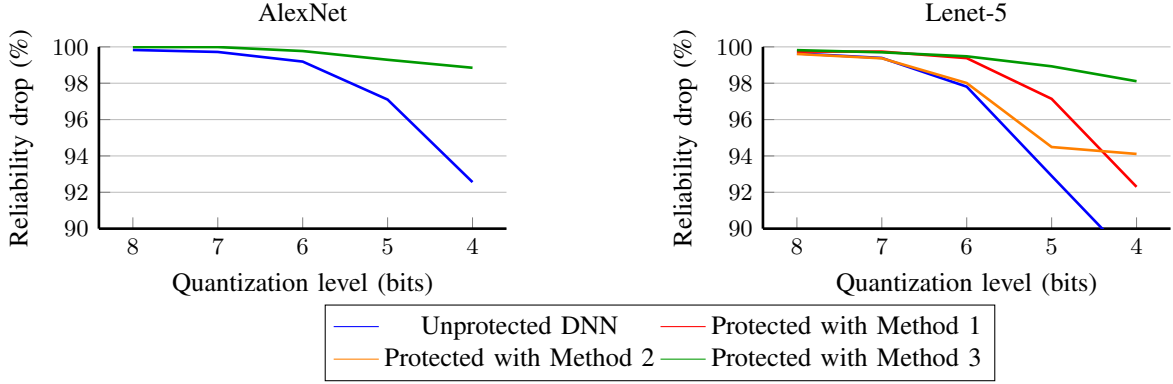
\begin{figure*}[h]
\begin{center}
\begin{subfigure}{.48\textwidth}
\begin{tikzpicture}
\begin{axis} [height=4cm, width=7cm, ymin=90, ymax=100, axis lines*=left,
              ymajorgrids=true, line width=1pt,
              x dir=reverse, xtick={8,7,6,5,4},
              enlarge x limits=0.1, tick label style={font=\small},
              xlabel=Quantization level (bits), ylabel={Reliability drop (\%)}, ylabel near ticks,
              title={AlexNet}]
\addplot[color=blue] coordinates {
	(8,99.83)
	(7,99.72)
	(6,99.19)
        (5,97.1)
	(4,92.56)
};
\addplot[color=green!60!black] coordinates {
	(8,99.99)
	(7,99.99)
	(6,99.77)
        (5,99.29)
	(4,98.85)
};
\end{axis}
\end{tikzpicture}
\end{subfigure}
\begin{subfigure}{.48\textwidth}
\begin{tikzpicture}
\begin{axis} [height=4cm, width=7cm, ymin=90, ymax=100, axis lines*=left,
              ymajorgrids=true, line width=1pt,
              x dir=reverse, xtick={8,7,6,5,4},
              enlarge x limits=0.1, tick label style={font=\small},
              xlabel=Quantization level (bits), ylabel={Reliability drop (\%)}, ylabel near ticks, title={Lenet-5},legend columns=2, legend entries={Unprotected DNN,Protected with Method 1, Protected with Method 2, Protected with Method 3},
              legend to name=methods]]
\addplot[color=blue] coordinates {
	(8,99.64)
	(7,99.39)
	(6,97.81)
        (5,92.90)
	(4,88.03)
};
\addplot[color=red] coordinates {
	(8,99.74)
	(7,99.74)
	(6,99.38)
        (5,97.14)
	(4,92.3)
};
\addplot[color=orange] coordinates {
	(8,99.61)
	(7,99.37)
	(6,98.01)
        (5,94.49)
	(4,94.11)
};
\addplot[color=green!60!black] coordinates {
	(8,99.82)
	(7,99.7)
	(6,99.48)
        (5,98.93)
	(4,98.11)
};
\end{axis}
\end{tikzpicture}
\end{subfigure}
\ref{methods}
\end{center}
\caption{Model-level reports of reliability drop (\%) based on different quantization degrees for AlexNet (left) and LeNet-5 (right)}
\label{re-improv}
\end{figure*}
\textbf{Step 4-B: Fault mitigation.}
Analyzing the output values of the network's intermediary layers post-training reveals identifiable upper and lower bounds for the neuron's output values. Leveraging this characteristic, we can ensure that any out-of-range outputs are reassigned to the respective upper or lower-bound values. This approach can be effectively implemented using specialized hardware units, as outlined below.

\textit{Out-range Error Detection}:
If the neuron's output value exceeds the predetermined upper or lower bound, it indicates a fault in the neuron's input values. To address this, a comparison is made between the neuron's output value and the two pre-established threshold values. For effective error detection, this paper introduces the following strategy.

For each layer, we store two values of upper bound and lower bound as the reference threshold for the out-ranged values. The output of the MAC (Multiply-Accumulate) unit is compared with the threshold values using two subtractors (negative values indicate that the output is beyond the threshold). The result of this comparison defines the final output (Fig. \ref{mit}). The general overhead of this mitigation technique is two stored values for each layer, and two subtractors to compare the MAC output value with the range threshold values and provide the select signal for the MUX to make the decision.

Three variations of this protection technique were implemented in the software to provide users with insights into the reliability enhancements this framework offers:
\begin{enumerate}
\item \textbf{Method 1:} When out-of-range value is detected it is replaced by the lower bound (min value).
\item \textbf{Method 2:} When out-of-range value is detected it is replaced by the upper bound (max value).
\item \textbf{Method 3:} When out-of-range value is detected it is replaced by either lower or upper bound depending on the sign of the MAC output.
\end{enumerate}

This protection technique is designed for easy replacement with any other protection methods (i.e. FT-ClipAct \cite{hoang2020ft}) within this framework toolchain without compromising the overall versatility of the framework.

\textbf{Step 5: Hardware generation.}

\begin{table*}[h]
\caption{Lenet-5 layer-level reports of fault criticality (\%) based on FI for different quantized networks}
\label{ax-accuracy-t}
\resizebox{18cm}{!}{
\begin{tabular}{|c"ccccc|ccccc|ccccc|ccccc|}
\hline
\begin{tabular}[c]{@{}c@{}}\% of critical\\ faults\end{tabular} & \multicolumn{5}{c|}{Unprotected}                                                                                          & \multicolumn{5}{c|}{Protected with Method 1}                                                                                                  & \multicolumn{5}{c|}{Protected with Method 2}                                                                                                  & \multicolumn{5}{c|}{Protected with Method 3}                                                                                              \\ \hline
Lenet-5                                                         & \multicolumn{1}{c|}{8 bit} & \multicolumn{1}{c|}{7 bit} & \multicolumn{1}{c|}{6 bit} & \multicolumn{1}{c|}{5 bit} & 4 bit & \multicolumn{1}{c|}{8 bit} & \multicolumn{1}{c|}{7 bit} & \multicolumn{1}{c|}{6 bit} & \multicolumn{1}{c|}{5 bit} & 4 bit & \multicolumn{1}{c|}{8 bit} & \multicolumn{1}{c|}{7 bit} & \multicolumn{1}{c|}{6 bit} & \multicolumn{1}{c|}{5 bit} & 4 bit & \multicolumn{1}{c|}{8 bit} & \multicolumn{1}{c|}{7 bit} & \multicolumn{1}{c|}{6 bit} & \multicolumn{1}{c|}{5 bit} & 4 bit \\ \thickhline
conv1                                                            & \multicolumn{1}{c|}{0.31}  & \multicolumn{1}{c|}{0.52}  & \multicolumn{1}{c|}{1.37}  & \multicolumn{1}{c|}{3.27}  & 9.12  & \multicolumn{1}{c|}{0.01}  & \multicolumn{1}{c|}{0}     & \multicolumn{1}{c|}{0.49}  & \multicolumn{1}{c|}{2.82}  & 9.06  & \multicolumn{1}{c|}{0.3}   & \multicolumn{1}{c|}{0.6}   & \multicolumn{1}{c|}{1.45}  & \multicolumn{1}{c|}{1.69}  & 3.76  & \multicolumn{1}{c|}{0}     & \multicolumn{1}{c|}{0}     & \multicolumn{1}{c|}{0.37}  & \multicolumn{1}{c|}{1.46}  & 3.49  \\ \hline
conv2                                                            & \multicolumn{1}{c|}{0.29}  & \multicolumn{1}{c|}{0.46}  & \multicolumn{1}{c|}{1.33}  & \multicolumn{1}{c|}{3.62}  & 9.38  & \multicolumn{1}{c|}{0.07}  & \multicolumn{1}{c|}{0.08}  & \multicolumn{1}{c|}{0.84}  & \multicolumn{1}{c|}{3.42}  & 8.49  & \multicolumn{1}{c|}{0.21}  & \multicolumn{1}{c|}{0.57}  & \multicolumn{1}{c|}{1.27}  & \multicolumn{1}{c|}{2.45}  & 4.71  & \multicolumn{1}{c|}{0.07}  & \multicolumn{1}{c|}{0.08}  & \multicolumn{1}{c|}{0.51}  & \multicolumn{1}{c|}{1.71}  & 4.08  \\ \hline
fc1                                                             & \multicolumn{1}{c|}{1.67}  & \multicolumn{1}{c|}{2.03}  & \multicolumn{1}{c|}{5.65}  & \multicolumn{1}{c|}{14.88} & 21.15 & \multicolumn{1}{c|}{1.04}  & \multicolumn{1}{c|}{0.9}   & \multicolumn{1}{c|}{2.14}  & \multicolumn{1}{c|}{6.67}  & 11.21 & \multicolumn{1}{c|}{1.72}  & \multicolumn{1}{c|}{1.78}  & \multicolumn{1}{c|}{4.91}  & \multicolumn{1}{c|}{9.23}  & 11.13 & \multicolumn{1}{c|}{0.82}  & \multicolumn{1}{c|}{1.18}  & \multicolumn{1}{c|}{1.82}  & \multicolumn{1}{c|}{3.2}   & 4.53  \\ \hline
fc2                                                             & \multicolumn{1}{c|}{1.6}   & \multicolumn{1}{c|}{2.41}  & \multicolumn{1}{c|}{5.88}  & \multicolumn{1}{c|}{16.31} & 25.5  & \multicolumn{1}{c|}{1.24}  & \multicolumn{1}{c|}{1.23}  & \multicolumn{1}{c|}{1.98}  & \multicolumn{1}{c|}{4.79}  & 13.68 & \multicolumn{1}{c|}{1.59}  & \multicolumn{1}{c|}{2.22}  & \multicolumn{1}{c|}{5.94}  & \multicolumn{1}{c|}{17.42} & 19.41 & \multicolumn{1}{c|}{0.97}  & \multicolumn{1}{c|}{1.26}  & \multicolumn{1}{c|}{2.24}  & \multicolumn{1}{c|}{3.09}  & 5.07  \\ \hline
\end{tabular}}
\end{table*}

\begin{table*}[]
\centering
\caption{AlexNet layer-level reports of fault criticality (\%) based on FI for different quantized networks}
\label{re_improv1}
\begin{tabular}{|c"ccccc|ccccc|}
\hline
\begin{tabular}[c]{@{}c@{}}\% of critical\\ faults \end{tabular} & \multicolumn{5}{c|}{Unprotected}                                                                                          & \multicolumn{5}{c|}{Protected with Method 3}                                                                            \\ \hline
AlexNet                      & \multicolumn{1}{c|}{8 bit} & \multicolumn{1}{c|}{7 bit} & \multicolumn{1}{c|}{6 bit} & \multicolumn{1}{c|}{5 bit} & 4 bit & \multicolumn{1}{c|}{8 bit} & \multicolumn{1}{c|}{7 bit} & \multicolumn{1}{c|}{6 bit} & \multicolumn{1}{c|}{5 bit} & 4 bit \\ \thickhline
conv1                         & \multicolumn{1}{c|}{0.5}   & \multicolumn{1}{c|}{0.79}  & \multicolumn{1}{c|}{1.76}  & \multicolumn{1}{c|}{4.03}  & 8.81  & \multicolumn{1}{c|}{0.05}  & \multicolumn{1}{c|}{0.06}  & \multicolumn{1}{c|}{0.52}  & \multicolumn{1}{c|}{1.87}  & 3.56  \\ \hline
conv2                         & \multicolumn{1}{c|}{0.58}  & \multicolumn{1}{c|}{1.05}  & \multicolumn{1}{c|}{1.35}  & \multicolumn{1}{c|}{1.66}  & 4.11  & \multicolumn{1}{c|}{0.03}  & \multicolumn{1}{c|}{0.03}  & \multicolumn{1}{c|}{1.035} & \multicolumn{1}{c|}{1.39}  & 3.31  \\ \hline
conv3                        & \multicolumn{1}{c|}{1.46}  & \multicolumn{1}{c|}{1.47}  & \multicolumn{1}{c|}{5.11}  & \multicolumn{1}{c|}{11.48} & 23.91 & \multicolumn{1}{c|}{0.07}  & \multicolumn{1}{c|}{0.08}  & \multicolumn{1}{c|}{1.14}  & \multicolumn{1}{c|}{1.29}  & 4.38  \\ \hline
conv4                        & \multicolumn{1}{c|}{0.99}  & \multicolumn{1}{c|}{1.63}  & \multicolumn{1}{c|}{2.46}  & \multicolumn{1}{c|}{7.13}  & 14.26 & \multicolumn{1}{c|}{0.03}  & \multicolumn{1}{c|}{0.04}  & \multicolumn{1}{c|}{1.30}  & \multicolumn{1}{c|}{4.13}  & 5.17  \\ \hline
conv5                        & \multicolumn{1}{c|}{0.90}  & \multicolumn{1}{c|}{2.10}  & \multicolumn{1}{c|}{3.69}  & \multicolumn{1}{c|}{7.82}  & 14.31 & \multicolumn{1}{c|}{0.04}  & \multicolumn{1}{c|}{0.09}  & \multicolumn{1}{c|}{1.61}  & \multicolumn{1}{c|}{3.44}  & 5.17  \\ \hline
fc1                          & \multicolumn{1}{c|}{3.02}  & \multicolumn{1}{c|}{4.95}  & \multicolumn{1}{c|}{8.15}  & \multicolumn{1}{c|}{16.38} & 31.19 & \multicolumn{1}{c|}{0.14}  & \multicolumn{1}{c|}{0.20}  & \multicolumn{1}{c|}{1.90}  & \multicolumn{1}{c|}{5.11}  & 8.66  \\ \hline
\end{tabular}
\end{table*}

At this step, a systolic-array-based QDNN accelerator for FPGA SoC is generated based on the parameters of the quantized network provided by \textbf{Step~4} to assess hardware utilization and requirements.

\begin{table}[h!]
\centering
\caption{SDC report for two unprotected Lenet-5 examples with different quantization levels}
\label{tab:sdc_metrics}
\begin{tabular}{lcc}
\hline
 Metric (\%)    &  16-bit  & 8-bit\\
\hline
 SDC-1          & 3.18   &  5.24\\
 SDC-5          & 28.04   & 37.26 \\
 SDC-10\%       & 14.30   & 17.65 \\
\hline
\end{tabular}
\end{table}

\begin{table*}[htp]
\centering
\caption{Model-level design space exploration results for Lenet-5 and AlexNet}
\label{dse}
\footnotesize
\begin{tabular}{|c|c|c|c|c|c|c|c|c|c|c|c|c|c|}
\hline
Network & BP & GIOPS & \multicolumn{3}{c|}{Resource utilization} & Accu- & Reliability & \multicolumn{3}{c|}{HW utilization (LUT)} & \multicolumn{3}{|c|}{Fault criticality improvement, \%} \\
\cline{4-6}\cline{9-14}
 & & & LUT & FF & DSP & racy, \% & improvement, \% & M1 & M2 & M3 & M1 & M2 & M3 \\
\hline
\multirow{6}{*}{Lenet-5}   & 16 & 0.058 & 5298 & 12,892 & 9  & 95.41 & ---   & 144
 & 144
 & 576
 & --- & --- & --- \\
                           & 8  & 0.079 & 3475 & 7003   & 9  & 94.02 & 64.33 & 72
 & 72
 & 288
 & 57.78 & 8.75  & 65.61\\
                           & 7  & ---   & ---  & ---    & ---& 93.93 & 67.95 &68
  & 68
 & 135
 & 71.24 & 14.95 & 67.74\\
                           & 6  & ---   & ---  & ---    & ---& 93.52 & 71.90 & 63
 & 63
 & 99 & 57.25 & 6.16  & 65.91\\
                           & 5  & ---   & ---  & ---    & ---& 92.49 & 81.17 & 68
 & 68
 & 81 & 36.30 & 31.34 & 66.86\\
                           & 4  & 0.087 & 2114 & 3865   & 9  & 89.65 & 81.17 & 36
 & 36
 & 63
  & 25.85 & 44.92 & 69.20\\
\hline
\multirow{6}{*}{AlexNet}   & 16 & 0.338 & 16,654 & 35,503 & 64 & --- & ---   & 1024
 & 1024
 & 2048 & --- & --- & --- \\
                           & 8  & 0.465 & 12,138 & 20,539 & 64 & 73.03 & 92.96 & 512
 & 512
 & 1024 & --- & --- & 94.27 \\
                           & 7  & ---   & ---    & ---    & ---& 72.26 & 89.79 & 480
 & 480
 & 960
 & --- & --- & 95.19 \\
                           & 6  & ---   & ---    & ---    & ---& 72.11 & 73.72 & 448
 & 448
 & 704
 & --- & --- & 58.59 \\
                           & 5  & ---   & ---    & ---    & ---& 70.69 & 66.32 & 480
 & 480
 & 576 & --- & --- & 54.13 \\
                           & 4  & 0.562 & 6428   & 10,067 & 64 & 69.15 & 78.07 &256
  & 256
 & 448 & --- & --- & 60.08 \\
\hline
\end{tabular}
\label{tab:metrics}
\end{table*}

The following tasks are executed at this step:

1. Network parameters are analyzed to determine the size of the systolic array, bit precision, and AXI bus bandwidth for data transfer. This analysis takes into account the number of kernels and feature map sizes. The goal is to optimize hardware accelerator performance for the generated network and improve overall efficiency.

2. The board is configured with the PYNQ bootable image. PYNQ provides Python and Jupyter Notebook support to AMD-Xilinx embedded devices. Included Python APIs allow to control both processing system and programmable logic (FPGA). PYNQ setup was selected to provide the users with a familiar interactive Python environment.

3. Network weights and biases are loaded on the board as NumPy array files. The network is described using a provided Python package that interfaces with the accelerator.

4. FPGA is configured from the Jupyter Notebook with the generated accelerator. Then, inference can be run using the provided input data.

\section{Experimental Results}

\subsection{Experimental setup}
Two networks are studied in this work: Lenet-5 and AlexNet. Lenet-5 is trained on the MNIST dataset, and AlexNet is trained on the CIFAR-10 dataset. Both networks are trained according to the \textbf{Step~1} methodology using quantization-aware training. Lenet-5 is trained using 16-bit INT data type, AlexNet is trained using 8-bit INT. For the study, different levels of quantization are applied in the \textbf{Step~2} using post-training quantization. 

Simulations are performed on 2 $\times$ Intel Xeon Gold 6148 2.40 GHz (40 cores, 80 threads per node) with 96 GB RAM. To speed up the simulation process, the framework supports multi-thread parallelism. 

To show the hardware characteristics of the output QDNN, studied networks are implemented on the Zynq UltraScale+ ZCU104 Evaluation Board (xczu7ev-ffvc1156-2-e).

\subsection{Fault simulator}
The fault simulator that is used in \textbf{Step~4} calculates the sufficient number of faults required for the reliability analysis. 
QDNNs generated by \textbf{Step~2} are validated by means of fault injection over the test set.

\emph{Random fault injection.} 
According to the adopted fault model, a random single bit-flip is injected into a random activation in a random layer of the network, and the whole test set is fed to the network to obtain the accuracy of the network. This process is repeated several times to reach an acceptable confidence level, which depends on the number of neurons and data representation bit length based on \cite{leveugle2009statistical}. This work provides an equation to reach 95\% confidence level and 1\% error margin. The framework adopts the formula presented in this work and provides a sufficient number of repetitions required for reliability analysis.

\subsection{Validation results}

The accuracy results for the quantized networks are reported in Table \ref{dse}. Further, fault injection is applied on each network automatically as part of the defined configuration of the framework, and reliability drop and fault criticality are reported in Fig. \ref{ax-accuracy} and Table \ref{ax-accuracy-t} for the Lenet-5 and in Fig. \ref{re_improv0} and Table \ref{re_improv1} for AlexNet. \textit{Reliability drop} is defined as the percentage of accuracy loss in the presence of the faults in the activations in a systolic-array-based simulation model of the network. \textit{Fault criticality} is defined as percentages of the faults that show a negative impact on the network accuracy and lead to misclassification. In Fig. \ref{ax-accuracy} and Table \ref{ax-accuracy-t}, the results for all versions of the proposed protection technique are documented for Lenet-5. Table \ref{re_improv1}, only the network protected with Method~3 is compared with the unprotected network for AlexNet, and in Fig. \ref{re_improv0} the reports the areliability drop without the protection techniques to show the impact of faults in activations, on different quantization level and layers of an AlexNet network. primarily due to space limitations within the paper.

From the previous works \cite{libano2021reduced}, it is evident that the reduction in memory size and quantization can lead to enhanced resilience and mitigate the impact of weight faults due to a reduced memory footprint. However, according to the presented charts, quantization may simultaneously heighten the network's vulnerability to faults in activations and logic. This is particularly crucial in lower precision networks, where even minor bit alterations can have significant ramifications. That is why reliability studies in the DNNs should be done for each QDNN to ensure the impact of quantization on the network's reliability.

Fig. \ref{ax-accuracy} shows that protection Method~3 is capable of improving the reliability of the network in the presence of a fault for more than 34.23\% in the worst case for Lenet-5. These numbers are calculated based on the following equation:
\[ \text{\% of Improvement} = \left( \frac{\text{New Value} - \text{Old Value}}{\text{Old Value}} \right) \times 100 \]
The same results are reported for AlexNet in Table \ref{dse}, which shows an improvement of more than 51.79\% in the worst case. Improvements in fault criticality for both networks at the model level are also reported in Table \ref{dse}, which demonstrates the positive impact of the protection technique on reducing the criticality of faults in both networks. These data also showcase the increasing fault criticality in different networks by increasing the level of quantization. Based on the results reported in Table \ref{dse}, protection Method~3, which shows the best results for improving reliability among all of the proposed protection techniques, introduces less than 10\% overhead compared to the LUTs required for the unprotected network implementation. Meanwhile, full protection of the network with TMR (Triple Module Redundancy) introduces more than 200\% hardware overhead. 

The fault injection procedure is performed for different quantizations and different versions of the proposed protection technique, and the accuracy drop, due to quantization and fault injection, is profiled. Further, in Table \ref{tab:sdc_metrics}, SDC metrics of two examples of quantized Lenet-5 are reported. It can be seen that these two networks are susceptible to injected faults. Specifically, the SDC-10\% and SDC-5 are very high: on average, about 3.18\% of the time the faulty inference misclassified the input in the 16-bit network and 5.24\% in the 8-bit network; furthermore, in 28.04\% cases for the 16-bit network and 37.26\% cases for the 8-bit network, the expected class is not even in the TOP-5 predictions. In addition, it can be observed that the 16-bit quantized network shows better performance in the presence of faults compared to the 8-bit network. In general, these results show that the DNNs used in this experiment are not suitable for a safety-critical application.

Hardware resource utilization and inference latency in GIOPS (Giga Integer Operations Per Second) for different quantization levels are reported in Table \ref{tab:metrics} alongside accuracy, reliability improvement due to the quantization, and hardware overhead and fault criticality improvement for fault mitigation techniques. These results of model-level design space exploration are provided for the user to understand the trade-off between reliability, accuracy, and required computational resources.

\section{Conclusion}

This paper presents a comprehensive methodology for exploring and enabling a holistic assessment of the trilateral impact of quantization on model accuracy, activation fault reliability, and hardware efficiency. A fully automated framework is introduced that is capable of applying various quantization techniques, fault injection, and hardware implementation, thus enabling the measurement of crucial hardware parameters like area and latency. Moreover, this paper proposes a novel lightweight protection technique integrated within the framework to ensure the dependable deployment of the final systolic-array-based FPGA implementation. The experiments on established benchmarks demonstrate the analysis flow and the profound implications of quantization on reliability, hardware performance, and network accuracy, particularly concerning the transient faults in the network's activations.

\section{Acknowledgement}
This work was supported in part by the Estonian Research Council grant PUT PRG1467 "CRASHLESS“ and by Estonian-French PARROT project "EnTrustED".

\bibliographystyle{IEEEtran}
\bibliography{ref}

\end{document}